\documentclass[aps,pra,superscriptaddress,twocolumn]{revtex4}
\usepackage{graphicx}
\usepackage{amssymb}
\usepackage{subfigure}
\usepackage{amsmath}
\usepackage{amsfonts}
\usepackage{bm}
\usepackage{color}
\newcommand{\beq}{\begin{equation}}
\newcommand{\eeq}{\end{equation}}

\newcommand{\ket}[1]{\ensuremath{\left|{#1}\right\rangle}}
\newcommand{\bra}[1]{\ensuremath{\left\langle{#1}\right |}}
\newcommand{\scalprod}[2]{\ensuremath{\left\langle{#1}|{#2}\right\rangle}}

\newcommand{\oper}[1]{\mathbf{\mathsf{#1}}}

\newcommand{\spww}[1]{{\color{red}\small}}

\begin{document}


\title{Free-Space Entangled Quantum Carpets}

\author{Mariana R. Barros}
\affiliation{Instituto de F\'{\i}sica, Universidade Federal do Rio de
Janeiro, Caixa Postal 68528, Rio de Janeiro, RJ 21941-972, Brazil}
\email{mariana@if.ufrj.br}
\author{Andreas Ketterer}
\affiliation{Laboratoire Mat\'eriaux et Ph\'enom\`enes Quantiques, Universit\'e Paris Diderot, CNRS UMR 7162, 75013, Paris, France}
\affiliation{Naturwissenschaftlich-Technische Fakult\"at, Universit\"aŠt Siegen, Walter-Flex-Str. 3, 57068 Siegen, Germany}
\author{Osvaldo Jim\'enez Far{\'\i}as}
\affiliation{Centro Brasileiro de Pesquisas F\'{\i}sicas, Rua Dr. Xavier Sigaud 150, Rio de Janeiro, 22290-180 Rio de Janeiro, Brazil}
\affiliation{ICFO-Institut de Ciencies Fotoniques, The Barcelona Institute of Science and
Technology, 08860 Castelldefels (Barcelona), Spain}
\author{Stephen P. Walborn}
\affiliation{Instituto de F\'{\i}sica, Universidade Federal do Rio de
Janeiro, Caixa Postal 68528, Rio de Janeiro, RJ 21941-972, Brazil}

\begin{abstract}
The Talbot effect in quantum physics is known to produce intricate patterns in the probability distribution of a particle, known as ``quantum carpets", corresponding to the revival and replication of the initial wave function.   Recently, it was shown that one can encode a $D$-level qudit, in such a way that the Talbot effect can be used to process the $D$-dimensional quantum information [Far\'{\i}as et al, PRA (2015)].  Here we introduce a scheme to produce free-propagating ``entangled quantum carpets" with pairs of photons produced by spontaneous parametric down-conversion.  First we introduce an optical device that can be used to synthesize arbitrary superposition states of Talbot qudits.  Sending spatially entangled photon pairs through a pair of these devices  produces an entangled pair of qudits.  As an application, we show how the Talbot effect can be used to test a $D$-dimensional Bell inequality.  Numerical simulations show that violation of the Bell inequality depends strongly on the amount of spatial correlation in the initial two-photon state.  We briefly discuss how our optical scheme might be adapted to matter wave experiments.       
\end{abstract}



\maketitle
\section{Introduction}
\begin{figure}
\centering 
\label{fig:talbotcarpet} 
\includegraphics[width=8cm]{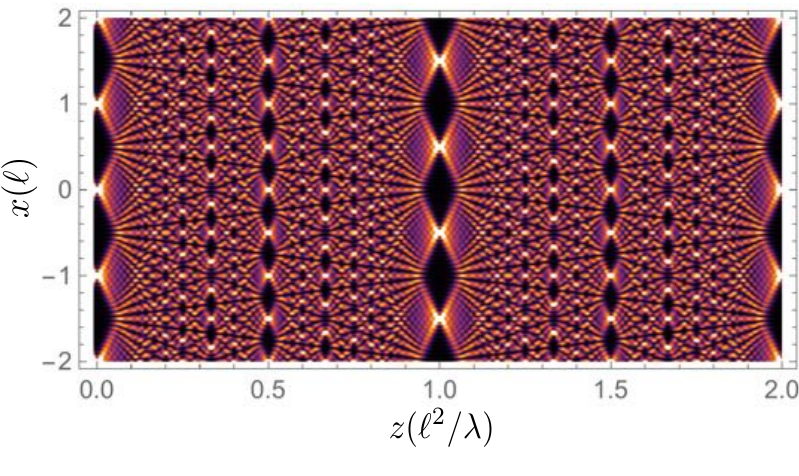}
\caption{{Quantum or Talbot carpet created by a plane wave propagating through a grating (period $\ell$) located at $z=0$. The transverse variable $x$ is given in units of $\ell$, and the  propagation distance $z$ is given in units of $\ell^2/\lambda$.}}
\end{figure}

Quantum carpets are elaborate patterns that appear in the probability density of a particle as a result of the self-interference of a wave packet, appearing due to its reflection from a pontential barrier, or to the periodic structure of the initial wave function \cite{berry96,berry01}.  This effect was first observed in classical optics by the British scientist Henry Fox Talbot in 1836 \cite{talbot1836} and subsequently explained by Lord Rayleigh in 1881 as a natural consequence of Fresnel diffraction \cite{rayleigh1881}.  Within the patterns of quantum or Talbot carpets,  one can observe the revival of the initial wave function, or even superpositions of displaced versions of the initial wave function.  The result is a intricate image of the probability density as it evolves, as shown in Fig. \ref{fig:talbotcarpet}.     
 
\par
The self-imaging effect appears in many physical systems such  as quantum wave packet revivals\cite{mondragon}, Bose Einstein condensates in optical lattices \cite{mostowski},  the magnetoelectric effect\cite{tokura}.  The Talbot effect can be used for routing and beam splitting in multimode waveguides \cite{soldano95}. 
\par
In the quantum optics regime, quantum Talbot carpets have been observed by sending twin photons through a periodic grating \cite{Luo09,Song11}, where here the characteristic Talbot length $z_T=\ell^2/\lambda$ is determined by the de Broglie wavelength of the two-photon wavepacket (here $\ell$ is the distance between slits in the grating and $\lambda$ the relevant wavelength).  Talbot carpets have been investigated experimentally in the classical and quantum regimes using multi-mode waveguides \cite{poem12,poem12b}.  Several of the above experiments have also been investigated using incoherent light sources \cite{Luo10,poem11}.   
\par
Recently, it was shown that  the Talbot effect can be used to process quantum information  \cite{farias15}.  In this scheme, propagation of fractions of the Talbot distance realize quantum gates on an appropriately chosen set of basis states composed of periodic wave functions--the ``Talbot basis".  Along with a phase gate, this scheme can be used to realize a universal gate set.  One interesting feature of this scheme is that it works for general dimension $D$, providing a scheme to encode and process $D$-dimensional quantum information.   
Higher-dimensional quantum systems are interesting for both fundamental as well as practical reasons.  The use of  $D$-dimensional ($D>2$) systems can allow for increased security and information transmission rate in quantum cryptography \cite{bechmann00a,bourennane01,walborn06a}, and higher-violation of Bell inequalities \cite{collins02}, which, in turn, can be used for device independent quantum key distribution (QKD) \cite{acin06} and generation of randomness \cite{acin10}.      
 \par
Here we introduce a method to produce free-space entangled quantum carpets and show how the natural evolution of the system, along with a simple phase element, could be used to realize quantum tasks, such as the violation of $D$-dimensional Bell inequalities.  Our proposal is made in the context of entangled photons, but the basic building blocks might also be realized in atom optics.  Our scheme differs from the previous work reported in Refs. \cite{Luo09,Song11} in that we produce two-photon states that are entangled in the Talbot basis. In addition, our scheme is designed for propagation in free-space rather than multimode waveguides, as in Refs. \cite{poem12,poem12b}. Another reason for using the phrase ``free-space'' is related to the fact that for communication tasks one must send the photons through some medium, e.g. fibers or free-space, whereas the latter is more applicable in the present scenario because fibers are usually not image preserving. 
 \par
This paper is organized as follows: in section \ref{sec:talbot} we review the Talbot effect and the encoding of $D$-dimensional quantum information using orthogonal periodic wave functions, which we refer to as ``Talbot Qudits".  We then show how simple quantum gates can be identified within the structure of the quantum carpet.  In section \ref{sec:EntCreation}, we show how arbitrary states of Talbot qudits can be produced, and we provide a scheme to produce arbitrary entangled quantum carpets in free space using spatially entangled twin photons.   The maximum dimension that can be encoded in the two photon state, under reasonable conditions, is also considered.  As an application, in section \ref{sec:bell} we investigate the violation of $D$-dimensional Bell inequalities via the Talbot effect.  Here the measurement bases are chosen using a spatially dependent phase element, such as a spatial light modulator, for example.  We provide numerical results taking into account relevant experimental parameters.  

\section{ D-dimensional quantum systems via the Optical Talbot Effect}
\label{sec:talbot}
 In the idealized version of the Talbot Effect, a plane wave of wavelength $\lambda$ propagates in the $z$ direction, and falls incident on a periodic grating that is located in the transverse plane at $z=0$. Let us assume that grating has period $\ell$, and that the central slit is offset from the $x=0$ origin by $x_d$. For simplicity we consider only one spatial dimension.  As long as the near field approximation is valid, the periodicity of the system is reflected not only in the transverse direction but also in the longitudinal direction.  At $z=0$ the wave function takes the form 
\begin{eqnarray}
\psi_d (x, z=0)= \sum_{n=-\infty}^{\infty}A_{n}e^{i n (x-x_d)k_{\ell}},
\label{eq:grating}
\end{eqnarray}
 where $A_n$ are the Fourier coefficients of the periodic grating. For simplicity, one can consider a grating made of arbitrarily-shaped slits of width $a$ that are  periodically separated by a distance $\ell$ (such that $\ell > a$), and then $k_\ell=2\pi/\ell$.  
In the paraxial approximation, propagation from $z=0$ to longitudinal position $z$ is given by \cite{goodman96}
\begin{equation}
\psi_d(x,z) = e^{i k z} \int dx^\prime \psi_d(x^\prime,0) e^{i k |x-x^\prime|^2/2z}. 
\label{eq:propgen}
\end{equation}  
Using the wave function \eqref{eq:grating} in \eqref{eq:propgen}, it is straightforward to arrive at
\begin{eqnarray}
\psi_d (x, z) = \sum_{n=-\infty}^{\infty}A_{n}e^{2 \pi i n (x-x_d)/\ell} e^{-i \pi n^2 z/z_T},
\label{eq:gratingz}
\end{eqnarray}
where $z_T = \ell^2/\lambda$ is the so-called Talbot length, which characterizes the longitudinal periodicity, as we will now see.  
For propagation distances satisfying $z^\prime=2 (s+q/r) z_T$ where $m$, $q$ and $r$ are coprime integers, one can show that \cite{case09}
\begin{equation}
\psi_d(x,z^\prime) = \sum_{j=0}^{r-1} a_j \psi_d \left(x-\frac{j}{r} \ell,0\right).  
\label{eq:talbotcarpet}
\end{equation}
The wave function $\psi(x,z^\prime)$ is a superposition of as many as $r$ versions of the original wave function, each offset by $j\ell/r$. The coefficients $a_j$ are given by the so-called Gauss sums, given by 
\begin{equation}
a_j = \frac{1}{r}\sum_{n=0}^{r-1} e^{-2 \pi i (n^2-jn)q/r}.
\label{eq:aj} 
\end{equation}
They depend on $r$ and $q$ but not on $s$, which reflects the longitudinal periodicity every $z=2z_T$.  

\par
If one considers single photons in a Talbot experiment, a structure of quantum levels can be defined using these displaced versions of the wavefunction.  Moreover, it has been shown that the Talbot effect can be used to perform quantum logic operations on these states \cite{farias15}.  
Let us choose the initial offset $x_d=d \ell/D$, where $d=\{0,\ldots,D-1\}$.  Then we can define a $D$-dimensional orthonormal basis of states $\ket{d_D}$ as
\begin{equation}
\scalprod{x}{d_D} \equiv \psi_d (x,0) =  \psi_{0}\left (x - \frac{d}{D} \ell \right). 
\label{qudits}
\end{equation}
We refer to this orthogonal basis of wavefunctions as the ``Talbot basis", and it will play the role of the computational basis.
Then, as suggested in Eq. \eqref{eq:talbotcarpet}, the Talbot effect can be viewed as a logic gate that transforms the basis elements $\ket{d_D}$ into superpositions of these states.  In fact, choosing $s=0$, $q=1$ and $r=c D$, where $c=1$ for odd $D$ and $c=2$ for even $D$, it is possible to show that
\begin{equation}
\psi_d (x, z=1/cD) =  \sum_{j=0}^{D-1} a_{cj} \psi_{0}\left (x - \frac{d}{D} \ell \right). 
\end{equation}

We can rewrite Eq. \eqref{eq:talbotcarpet} in the vector space notation as
\begin{equation}
\ket{\psi_d(z)}= \sum_{j=0}^{r-1} a_{cj}\ket{j}=\oper{T}_{cD}\ket{0}
\label{eq:TalbotGate}
\end{equation}
where we introduced the Talbot gate $\oper{T}_{cD}=\sum_j^{r-1}a_{cj} \oper{X}^j$ and $\oper{X}$ is the generalized Pauli operator in dimension $D$: $\oper{X}^j\ket{d}=\ket{(d+j)_{mod D}}$. 

\par In order to further manipulate the Talbot qudits defined in \eqref{qudits}, we introduce a phase gate
\begin{equation}
\label{eq:ztheta}
Z_{\vec{\theta}}= \sum^{D-1}_{d=0} e^{i\theta_{d}}\left|d_{D}\right\rangle\left\langle d_{D}\right|,
\end{equation}
where $\vec{\theta}=\{\theta_0,\dots,\theta_{D-1}\}$ are the phases applied to each Talbot basis state. The phase gate $Z_{\vec{\theta}}$ is diagonal in the position basis, and thus can be implemented using a spatial light modulator, or any other optical element that imprints the appropriate transverse phase on a single-photon field.  The Talbot gate along with the phase gate are enough to perform a universal set of single qudit  \cite{farias15}.   Combining sequential phases and free space propagation allows one to produce a complex spatial structure, a ``quantum carpet", where the output wave function corresponds to the desired quantum operation.  In the next section, we show how to produce entangled quantum carpets.   
   
\section{Entangled Talbot Carpets with photon pairs} 
\label{sec:EntCreation}
In this Section we introduce a method  to create $D$-dimensional entangled Talbot states, such as
\begin{align}
\ket{\text{ET}_D}= \sum_{d=0}^{D-1} C_d \ket{d_D}\ket{d_D},
\label{eq:TalbotEnt}
\end{align}
where the Talbot basis states $\ket{d_D}$ were defined in Eq. \eqref{qudits}.  
Our scheme will rely on producing spatially entangled photon pairs from spontaneous parametric down-conversion (SPDC).  We will then introduce an optical device that we call a  ``Talbot carpet synthesizer", which can be used to produce arbitrary superposition states in the Talbot basis.  We will then show that applying this device to the SPDC photon pairs allows one to produce entangled states of the form \eqref{eq:TalbotEnt}.  
\subsection{Spatial correlations}
\begin{figure}
\includegraphics[width=0.48\textwidth]{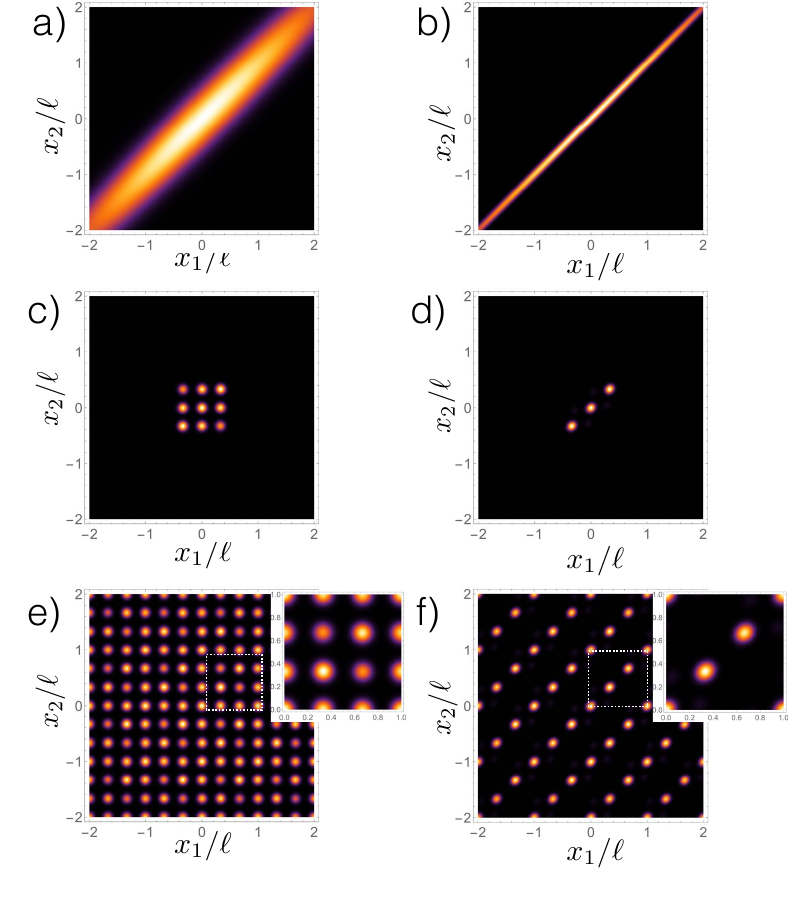}
\caption{(color online) Density plots showing different steps in the creation of entangled quantum carpets.  On all figures on the left side, we have reasonable initial spatial entanglement with of the wave function with $\kappa_+=9\ell$ and $\kappa_-=\ell$($R=0.975$), while on figures on the right we have considerable entanglement with $\kappa_+=9\ell$ and $\kappa_-=1/6\ell$($R=0.999$).  Figures a) and b) show plots of the inital spatially entangled state $\psi_{12}(x_1,x_2)$ with $\psi_+(x)$ and $\psi_-(x)$ chosen as Gaussians with widths $\kappa_+$ and $\kappa_-$.  Figures c) and d) show the state $\psi^{(D)}_{12}(x_1,x_2)$ after the spatially entangled photons pass through gaussian $D=3$-slit apertures with width $\delta=0.05\ell$ separated by $\ell$.  Figures e) and f) show the final states after passing through Gaussian gratings with slit width $\sigma=0.05\ell$ and distance $\ell_g=\ell$ for $D=3$.}
\label{fig:Carpet_Gen}
\end{figure}

Spatial correlations from SPDC have been studied in a number of contexts \cite{monken98a,howell04,neves05,walborn07c,pires09,peeters09,walborn10,edgar12,carvalho12,howland13,schneeloch15}.  
We start by recalling that the transverse state of two photons produced in SPDC can be expressed as follows \cite{monken98a,walborn10,schneeloch15}:
\begin{align}
\ket\Psi_{12}=\iint dx_1 dx_2 \Psi_{12}(x_1,x_2) \ket{x_1}\ket{x_2},
\label{eq:PDC}
\end{align}
where 
\begin{align}
\Psi_{12}(x_1,x_2)=\psi_+(x_1+x_2) \psi_-(x_1-x_2),
\label{eq:wavefct_1}
\end{align}
is the transverse wave function of the twin photons and $\ket{x_1} \ket{x_2}$ a two mode Fock state describing a single transverse mode of each photon. The functions $\psi_+(x_1+x_2)$ and $\psi_-(x_1-x_2)$ are determined by the pump field profile and the phase matching function of the SPDC process, respectively.  
In typical SPDC experiments, one can employ the ``double gaussian approximation" \cite{walborn10,schneeloch15}, in which both $\psi_+$ and $\psi_-$ are described by  Gaussian functions with widths $\kappa_+$ and $\kappa_-$, respectively.   The correlation coefficient $R = (\kappa_+^2 - \kappa_-^2)/ (\kappa_+^2 + \kappa_-^2)$ describes the spatial correlation between the photons.  For $R=\pm1$, the photons are perfectly correlated:  $x_1=\pm x_2$.  In the near-field it is typical that $\kappa_- << \kappa_+$, and $\kappa_-$ is proportional to the transverse correlation length of the photon pair in the source plane \cite{schneeloch15}. Plots of the wave function (\ref{eq:wavefct_1}) are presented in Fig.~\ref{fig:Carpet_Gen} a) and b) for different values of $R$. 
\par
Our approach to generate entangled Talbot carpets will take advantage of the spatial correlations of photon pairs generated from SPDC.  First, we will introduce an apparatus that can be used to produce arbitrary superposition states of Talbot qudits.  Then, we will show how this can be used to produce entangled Talbot qudits.
\begin{figure}
\includegraphics[width=6cm]{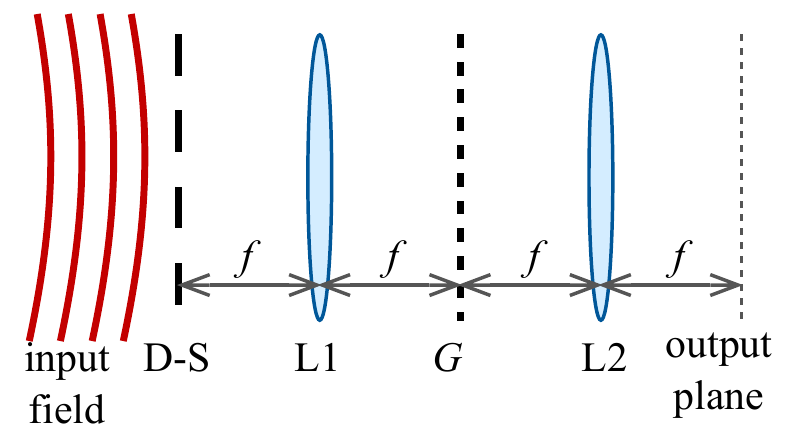}
\caption{(color online) Proposal to produce arbitrary Talbot states.  A nearly plane wave is incident on a $D$-slit aperture (D-S) creating the state $\psi_{D}(x)$, followed by the transformation shown in Eq. (\ref{eq:psi_convo}). Lenses (L1,L2) implement two Fourier transforms and the grating (G) is placed in the Fourier plane in between (L1) and (L2). (SC) denotes the detection screen.  }
\label{fig:scheme_1}
\end{figure}
\subsection{Talbot carpet synthesizer}
Here we present a device that converts a monochromatic single-photon field into an arbitrary superposition of Talbot basis states.  The scheme is illustrated in Fig.~\ref{fig:scheme_1}.  
We consider a monochromatic plane wave incident on a Young-like aperture consisting of $D$ slits of width $\delta$ separated by a distance $s$ ($s> \delta$).    
Assuming that the incident wave function $\mathcal{E}_{in}(x)$ is approximately constant over each slit, directly after the aperture the outgoing wave function is  given by 
\begin{align}
S_{D}(x)= \sum_{d=0}^{D-1} c_d S (x-d s),
\label{eq:psi_Dslit}
\end{align}
where $S(x)$ defines the transmission profile of each of the identical slits, and the complex amplitudes are given by $c_d \propto \mathcal{E}_{in}(ds)$ and $ \sum_{d=0}^{D-1} |c_d|^2=1$.  The state \eqref{eq:psi_Dslit} is an arbitrary superposition state of the $D$ states describing passage through each of the $D$ slits.  In principle, one can control the amplitudes $c_d$ by controlling the amplitude and phase of the input field before the slit aperture.   
 In order to transform Eq.~(\ref{eq:psi_Dslit}) into a $D$-dimensional Talbot state, we apply a lens (L1) to implement a Fourier transform, and place a grating $G$ at the Fourier plane of the lens.   A second lens (L2) performs a second Fourier transform.  For simplicity we assume that each lens has focal length $f$. The wave function of the state resulting from this transformation is given by:
\begin{align}
\psi_{out}(x)=\mathcal F\{ G(x) \cdot \mathcal F \{S_D(x)\}\}=\left(\mathcal F\{G\} \ast S_D\right) (x),  
\label{eq:psi_convo}
\end{align}
where $\mathcal{F}$ is the Fourier transform and ``*" means convolution.  In the second equality in Eq.~(\ref{eq:psi_convo}) we used the convolution theorem for Fourier transforms.   We see that the output field at the Fourier plane of lens L2 is the convolution of the Fourier transform of the grating $G$ with the original $D$-slit state $S_D$.    
Writing the grating transmission function as $G(x)=\sum_m A_m \exp({2\pi i m x/\ell_g})$ (see Eq. \eqref{eq:grating}), the Fourier transform of $G(x)$ is given by 
\begin{align}
\mathcal F\{G\}(x^\prime)= \sum_{m=-\infty}^{+\infty} A_m \delta \left (\frac{m}{\ell_g} - \frac{x^\prime}{ f \lambda}\right).
\label{eq:G}
\end{align}
Using equations \eqref{eq:psi_Dslit} and \eqref{eq:G} in \eqref{eq:psi_convo}, we have
\begin{equation}
\psi_{out}(x)=  \sum\limits_{d=0}^{D-1} c_d T_d(x), 
\label{eq:psi_talbot}
\end{equation}
where 
\begin{equation}
T_d(x)=   \sum\limits_{m=-\infty}^{m=+\infty}A_m  S\left(x-d s -\frac{mf \lambda}{\ell_g} \right )
\label{eq:T}
\end{equation}
describes the transmission function of an  ``effective grating" consisting of an infinite periodic comb of slit functions $S$, each centered at $ds - mf\lambda\ell_g$.  The period of the effective grating is $\ell =  f \lambda/\ell_g$. The function $T_d(x)$ can then be chosen as the wave function of the Talbot basis states:  $T_d(x) = \langle x\ket{d_D}$. The state at the output plane \eqref{eq:psi_talbot} is thus an arbitrary superposition of Talbot states, controlled by the amplitudes $c_d$.   
\par
\begin{figure}
\includegraphics[width=0.475\textwidth]{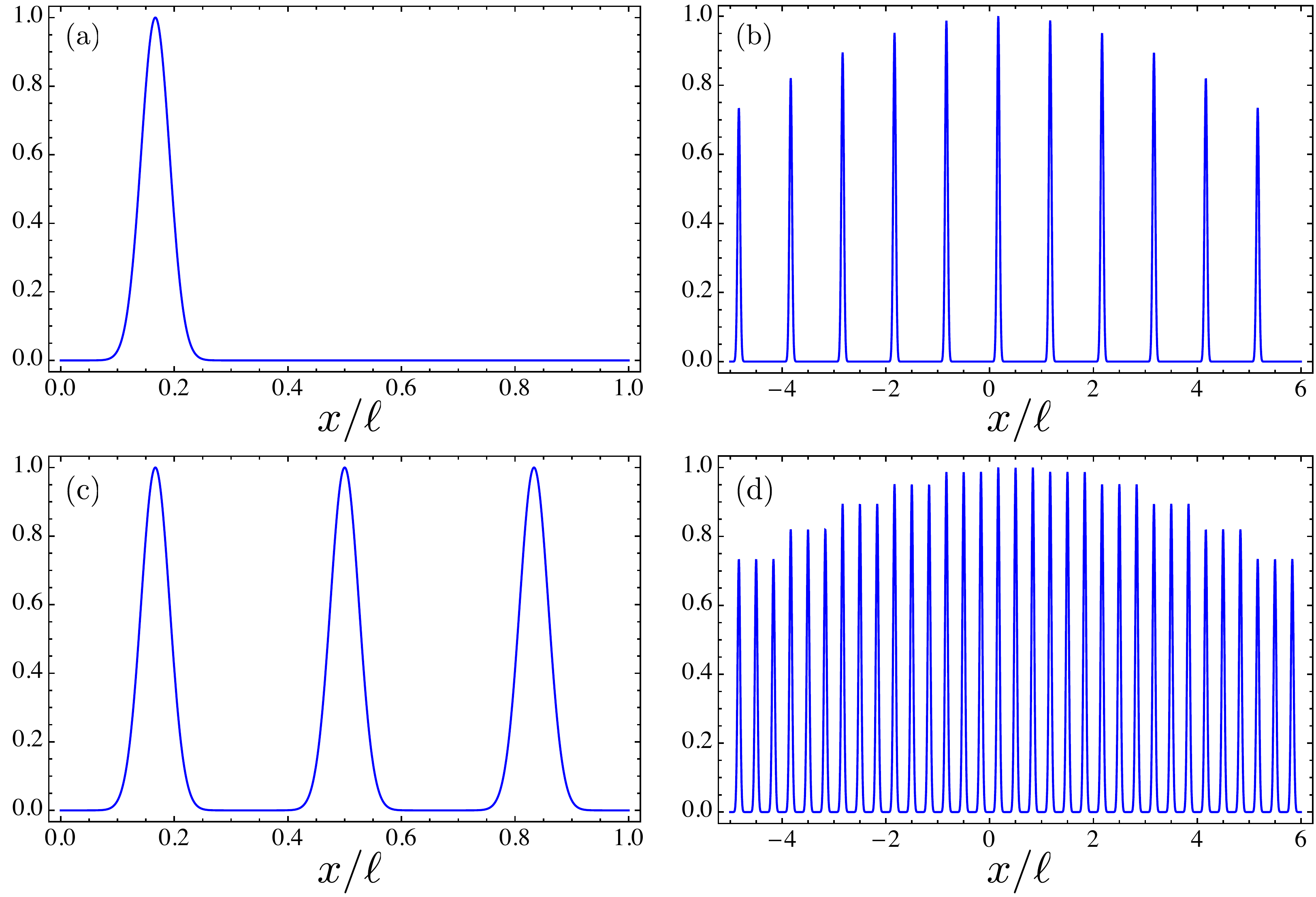}
\caption{(color online) Plots of the wave function $\psi_{D=3}(x)$ (left column) and its transformed version $\tilde\psi_{D=3}(x)$ (right column), with a Gaussian slit transmission profile $S(x)$ with width $\Delta=0.025s$, a Gaussian grating transmission function $G(x)$ with width $\kappa=0.025 s$ and slit distance $\ell_g=s$.  Shown are two examples: amplitudes $c_0=1$, $c_1=0$, $c_2=0$ in (a,b) and $c_n=1/\sqrt D$ in (c,d).}
\label{fig:plot_Dslit}
\end{figure}

 As example, we treat the case where the transmission profile $S(x)$ of the slits in the $D$-slit aperture is a Gaussian function with width $\Delta$, and the grating transmission function is given by a Gaussian comb with Fourier coefficients $A_m=e^{-(2 \pi m \sigma)^2/(2\ell_g^2)}$, where $\sigma$ is the width of the Gaussian spikes and $\ell_g$ the distance between them. In Fig.~\ref{fig:plot_Dslit}, we plot examples of the wave functions $S(x)$ and $\tilde\psi_D(x)$ for this case with  $D=3$. We chose Gaussian functions for computational simplicity, however any function can be used. 

\subsection{Entangled Talbot qudits}
\begin{figure}
\includegraphics[width=0.4\textwidth]{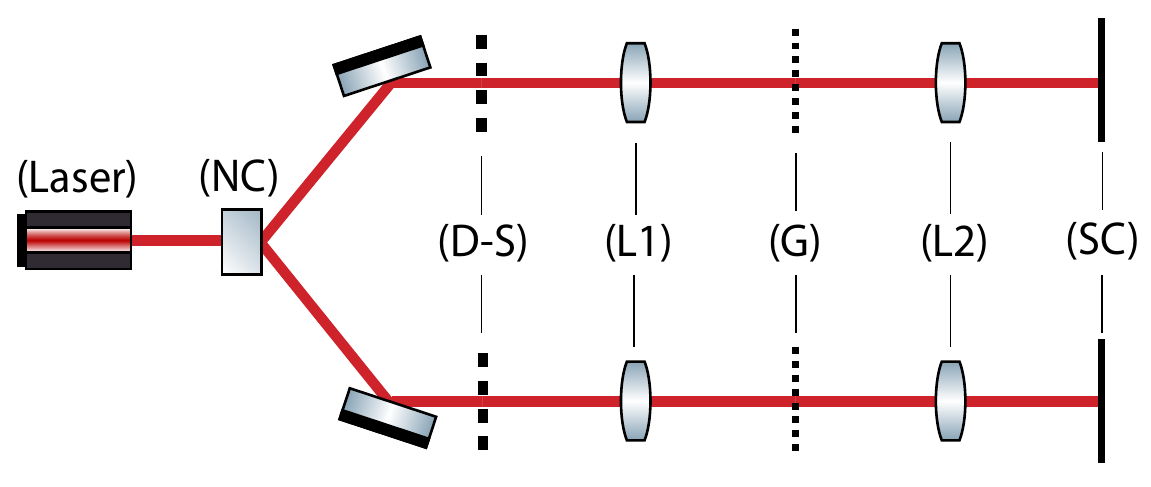}
\caption{(color online) Experimental proposal: A laser is incident on a non-linear crystal (NC) producing two down-converted photons in the state $\ket\psi_{\text{SPDC}}$ and subsequently pass through individual $D$-slits apertures (D-S) creating the state $\ket\psi_{\text{SPDC,g}}$. The following  elements implement on each photon the same transformation as in Fig. \ref{fig:scheme_1}. (SC) denotes the detection screen.   }
\label{fig:scheme_2}
\end{figure}
We now turn to the question of how to produce a $D$-dimensional entangled Talbot state from SPDC. The corresponding experimental scheme for this task is depicted in Fig.~\ref{fig:scheme_2}.  The basic idea is to use the Talbot state synthesizer from last section, one for each photon, and engineer the spatial correlations to control the entanglement in the Talbot basis.   
As shown in the figure, we send the down-converted photons through the same optical elements as in Fig. \ref{fig:scheme_1}.  First, after the identical $D$-slit apertures the wave function (\ref{eq:wavefct_1}) becomes:
\begin{align}
\psi^{(D)}_{12}(x_1,x_2)=S_{D}(x_1) S_{D}(x_2)  \psi_+(x_1+x_2) \psi_-(x_1-x_2),
\label{eq:SPDCgrating}
\end{align}
where $S_{D}$ is the transmission functions of the $D$-slit apertures given in Eq. \eqref{eq:psi_Dslit}. 
It is well known that under proper conditions the two-photon wave function \eqref{eq:SPDCgrating} describes entangled ``Young Qudits" \cite{neves05,peeters09,carvalho12}.  In particular, when the correlation $R \approx 1$, the photons can be made to almost always pass through the same slit, giving rise to a two-photon interference pattern that is described by the two-photon de Broglie wavelength \cite{fonseca99b,peeters09}.   
To illustrate this, plots of the wave function (\ref{eq:SPDCgrating}) with $S(x)$ given by a Gaussian, are presented in \ref{fig:Carpet_Gen} c) and d) for the same values of $R$ used in figure \ref{fig:Carpet_Gen} a) and b), respectively. 

\par
After the $D$-slits we apply to both photons the same sequence of operations:  a Fourier transform, a grating operation and another Fourier transform (see Eq. (\ref{eq:psi_convo})), giving:
\begin{equation}
\Psi(x_1,x_2)= [\mathcal{F}\{G\}(x_1) \mathcal{F}\{G\}(x_2)] * \psi^{(D)}_{12}(x_1,x_2).
\end{equation}
Using Eqs.  \eqref{eq:G}  and \eqref{eq:SPDCgrating}, 
\begin{align}
\Psi(x_1,x_2)=& \sum_{d_1,d_2} \sum_{m_1=-\infty}^{\infty} \sum_{m_2=-\infty}^{\infty} A_{1,m_1} A_{2,m_2} \nonumber \\
& \times S\left(x_1-\frac{m_1 f \lambda}{\ell_g}-d_1s\right)S\left(x_2-\frac{m_2 f \lambda}{\ell_g}-d_2s\right) \nonumber \\
& \times \psi_{\text{12}}\left(x_1-\frac{m_1 f \lambda} {\ell_g}, x_2-\frac{m_2 f \lambda}{\ell_g}\right).
\label{eq:PsiFinal}
\end{align}

If the width $\delta$ of the slit functions $S$ is small compared to the widths $\kappa_\pm$ describing the spatial correlations of the two-photon state, we can approximate
\begin{equation}
 \psi_{12}\left(x_1-\frac{m_1 f \lambda} {\ell_g}, x_2-\frac{m_2 f \lambda}{\ell_g}\right) \approx \psi_+(d_1s + d_2s) \psi_-(d_1s - d_2s).
\end{equation}
Using Eq. \eqref{eq:T} we can rewrite Eq. \eqref{eq:PsiFinal}  as
\begin{equation}
\Psi(x_1,x_2)= \sum_{d_1,d_2} C_{d_1,d_2}T_{d_1}(x_1) T_{d_2}(x_2), 
\label{eq:PsiFinal2}
\end{equation}
where 
\begin{align}
C_{d_1,d_2} & =  \psi_+(d_1s + d_2s) \psi_-(d_1s - d_2 s) \nonumber \\
& = A \exp\left(-\frac{s^2}{4 \Delta_{+}^2}\left(d_1^2 + \frac{2 \Delta_{+}^2}{\Delta_{-}^2}d_1d_2 + d_2^2 \right ) \right),
\label{eq:Cdd}
\end{align}
and $A$ is a normalization factor, such that $\sum_{d_1,d_2} |C_{d_1,d_2}|^2=1$.  Here $1/\Delta_{\pm}^2 = 1/\kappa_+^2 \pm 1/\kappa_-^2$.  
\par
The final state \eqref{eq:PsiFinal2} is a non-local superposition of effective grating states for photons 1 and 2, an \emph{entangled quantum carpet}.  The entanglement depends explicitely on the coefficients $C_{d_1,d_2}$.  In the coefficient of the cross term  in \eqref{eq:Cdd} we identify the spatial correlation coefficient of the photon pair defined above:  $\Delta_+^2/\Delta_-^2 = -R$.  When the spatial correlation is large so that $\kappa_- << \kappa_+$ and $R \approx 1$, we have $C_{d_1,d_2} \approx \exp[-s^2(d_1-d_2)^2/4 \Delta_+^2]$. To ensure a large amount of entanglement in the $D\times D$ dimensional two photon Talbot state requires $s^2/4\Delta_+^2 >> 1$, resulting in $\kappa_-^2 << s^2$. In other words, to produce entangled Talbot states we need a transverse correlation length at the SPDC source that is smaller than the distance between slits in the $D$-slit apparatus.

As example, we plot the wave function $\Psi(x_1,x_2)$ for the case $D=3$ in Fig. \ref{fig:Carpet_Gen} e) and f) using the same Gaussian slit profile and grating transmission functions as in previous examples.  One can see the effect of the initial spatial entanglement on the entangled quantum carpet.  In Fig. \ref{fig:Carpet_Gen} a), there is little initial spatial entanglement, which can be seen by the illumination of all 9 slit combinations in Fig. \ref{fig:Carpet_Gen} c),  giving rise to a less entangled quantum carpet in Fig. \ref{fig:Carpet_Gen} e).  On the other hand Fig. \ref{fig:Carpet_Gen} b)  shows the wave function produced by SPDC for a reasonable amount of spatial entanglement, which guarantees that the photons pass through the same slit in the $D$-slit apparatus, as shown in Fig. \ref{fig:Carpet_Gen}.  This gives rise to the entangled quantum carpet shown in Fig. \ref{fig:Carpet_Gen} f).  In the next section we show how these entangled quantum carpets can be used to violate $D$ dimensional Bell inequalities.     

\subsection{Physical constraints on Talbot qudits}
The resolution and size of the SLMs and CCD cameras (or detector arrays) that are used to manipulate and measure the spatial field of the photons set constraints on the dimension $D$ that one is able to encode. In the following, we will elaborate on these constraints and provide an example with realistic experimental parameters.

 Let us first assume that the used SLM and CCD camera (SLM/CCD) have a finite resolution but their size is infinitely extended in the transverse plane. In this case it is in principle possible to encode arbitrarily large Hilbert spaces in the field of the photon. One just needs to keep in mind that there will be a relation between the encoded dimension $D$ and the corresponding Talbot length $z_T$, given by: 
 \begin{align}
 z_T=\frac{\ell^2}{\lambda}=\frac{\rho^2 D^2}{\lambda},
\end{align}
where $\rho$ denotes the resolution of the SLM/CCD and $\lambda$ the corresponding wavelength of the photon. Note that the length $\ell=\rho D$ refers to the distance between those peaks of the periodic wave function which correspond to the same basis wave function $\scalprod{x}{d_D}$. Hence, the Talbot length increases quadratically with the the encoded dimension $D$. Note that the propagation distance required to implement the Talbot gate, given by $z_T/(gD)$, where $g=1,2$ for $D$ even and odd, respectively, increases linearly with $D$. However, as long as there is no restriction on the transverse width of the SLM/CCD there is in principle no constrain on the size of the encoded dimension $D$.

Now we turn to the more realistic situation and assume that the SLM/CCD has a finite extension. 
 In order to explain this, we note that the use of the Talbot effect to manipulate $D$ dimensional quantum information relies on the paraxial approximation (see Sec.~\ref{sec:talbot}). Hence, any deviation from the latter will lead to a decrease in the fidelity of the initial  state with respect to itself after multiples of the Talbot propagation distance. One possible source of such deviations are non-perfect plane waves that illuminate only a finite number of slits of the respective grating (or SLM) that produces the Talbot qudits. Another source for deviations is the finite size of the optical elements used to manipulate or measure the photons, \textit{e.g.} the SLM or the CCD camera, which leads to a similar effect. Hence, given a certain number of illuminated slits (which is related to the size of the SLM/CCD given a fixed resolution $\rho$), one can propagate the Talbot qudits for a certain number of Talbot distances before their fidelity begins to decrease considerably. 

In \cite{farias15} it was shown that for about $100$ illuminated slits (or more), the fidelity of the Talbot qudits is fairly well preserved for propagation distances of several Talbot distances. For many entanglement-based purposes, such as QKD, it is not necessary to have free-propagation for more than one or two Talbot distances. Of course, the photons need to be sent from one place to another which can be done with optical imaging systems that produce images of the gratings at the desired regions. The single-photon fields can then be allowed to free-propagate the distances required to perform the relevant operations. 

The threshold of $100$ illuminated slits leads to a bound on the dimension one can encode into the spatial field of the photons. To show this, we divide the size of the SLM $s_{\text{SLM}}$ by the length $\ell=\rho D$ and set this fraction equal to 100. This leads to the following formula for the encoded dimension: 
\begin{align}
D=\frac{ s_{\text{SLM}}}{100 \rho}.
\end{align}
For a typical SLM we have a resolution of about $10\mu m$ per pixel and an overall  size of $1080\times1920$ pixels, leading to an estimated dimension $D=19$ (If we use the larger side of the SLM to manipulate the Talbot qudits). The finite size of the SLM in the second transverse dimension has little effect, since it is usually much larger than the field distribution in that direction which is typically about few millimeters in size.

A bound on the encoded dimension $D$ directly leads to a bound on the mutual information $I_{\text{AB}}=-\frac{1}{D}\sum_{k=0}^{D-1}\log_2{(1/D)}$ one can encode into a pair of photons and reliably manipulate without considerable loss of fidelity.  For instance, with $D=19$ one can send about $4.25$ random bits per photon pair in a QKD scheme (before any post-processing such as basis reconciliation, error correction, etc).  
\par
The limit given above considers that an entire SLM can be illuminated with constant intensity.  In the next section, we consider the violation of $D$-dimensional Bell inequalities as a benchmark for the entangled quantum carpets, providing numerical examples in section \ref{sec:bellexample} that take into account the finite entanglement of the two-photon state, as well as the finite widths of the pump laser beam and phase matching function in SPDC.   
\section{Violation of $D$-dimensional Bell's inequalities}
\label{sec:bell}
In order to evaluate the quality of the entangled quantum carpets as a function of the initial spatial entanglement of the down-converted photon pairs, and to show the utility of the quantum gates produced via the Talbot effect, we consider the violation of $D$-dimensional Bell inequalities. States violating these types inequalities can be used for QKD and other (semi-)device-independent quantum information protocols.

An interesting set of Bell-like inequalities are the those developed by Collins et al \cite{collins02}.  Alice and Bob can carry out two possible measurements, 1 or 2, giving result $A_{1}$ or $A_2$ ($B_1$ or $B_2$)  when Alice (Bob) performs measurement 1 or 2.  Each measurement may have $D$ possible outcomes: $A_{1}, A_{2}, B_{1}, B_{2} = 0, \ldots, D -1$. 
Local realistic distributions satisfy the inequality 
\begin{equation}
I_D = \sum\limits_{k=0}^{[D/2]-1} \left (1-\frac{2k}{D-1} \right ) J_k \leq 2,
\label{eq:ID}
\end{equation}
 where
\begin{align}
J_k =  & P(A_{1}= B_{1}+k) - P(A_{1}= B_{1}-k-1)   \nonumber \\
+ & P(B_{2}=A_{1}+k)  - P(B_{2}=A_{1}-k-1) \nonumber \\
+ & P(B_{1} = A_{2} +k+1) - P(B_{1} = A_{2} -k) \nonumber \\
+ & P(A_{2}=B_{2}+k) - P(A_{2}=B_{2}-k-1),
\end{align}
and
the joint probabilities $P$ are defined as
\begin{equation}
P(A_{a} = B_{b} + k) = \sum^{D-1}_{j=0} P(A_{a}=-j, B_{b}=j+ k\mod D),
\end{equation}
where we have adapted the correlation probabilities of Ref. \cite{collins02} to give $P=1$ when Alice and Bob obtain perfectly anti-correlated results. These inequalities were shown to be more resistant to noise \cite{collins02}.  Moreover, they provide an increase in the quantum mechanical violation as a function of $D$.  For $D=2$ we have $I_2 = 2\sqrt{2} \approx 2.828$, while for $D\longrightarrow \infty$, $I_\infty \approx 2.9681$.  

\par
Violation of the CGLMP inequality can be obtained by considering maximally entangled $D$-dimensional systems of the form \cite{collins02} 
\begin{equation}
\ket{\phi} = \frac{1}{\sqrt{D}} \sum\limits_{d=0}^{D-1} \ket{d_D}_A \ket{d_D}_B, 
\end{equation}
where Alice and Bob perform projective measurements onto bases defined by the eigenstates:
\begin{equation}
\left|f_\alpha \right\rangle = \frac{1}{\sqrt{D}} \sum_{d=0}^{D-1} e^{2 \pi i d (f + \alpha)/D} \left|d_{D}\right\rangle,
\label{eq:k}
\end{equation}

\begin{equation}
\left|f_\beta \right\rangle = \frac{1}{\sqrt{D}} \sum_{d=0}^{D-1} e^{2 \pi i d (-f + \beta)/D} \left|d_{D}\right\rangle,
\label{eq:l}
\end{equation}
with $\alpha$ and $\beta$ taking the following values for measurements 1 and 2:  $\alpha_{1} = 0$,  $\alpha_{2} = \frac{1}{2}$, $\beta_{1}= \frac{1}{4}$ and $\beta_{2}= -\frac{1}{4}$. 
In the next section we show that these projective measurments can be realized using the Talbot effect.
\subsection{Bell's inequality experiments with the Talbot Effect}
We consider that Alice and Bob are able to project onto the computational basis, defined by the states $\ket{d_D}$.  As discussed in Ref. \cite{farias15}, this can be performed using an array of single-photon detectors  or CCD cameras sensitive to single photons  \cite{edgar12}.   Thus, Alice and Bob can project in the bases defined by \eqref{eq:k} and \eqref{eq:l} if they have unitary operators $\oper{U}$ which map:
\begin{eqnarray}
\oper{U}_{A}\left|d_{\alpha}\right\rangle = e^{i g(d,\alpha)} \left|d_{D}\right\rangle \\ 
\oper{U}_{B}\left|d_{\beta}\right\rangle = e^{i h(d,\beta)} \left|d_{D}\right\rangle,
\end{eqnarray}
where $g(d,\alpha)$ and $h(d,\beta)$ are arbitrary functions. 
Through direct calculations one can see that the unitaries $\oper{U}_{A}$ and $\oper{U}_{B}$, that map these eigenstates into the computational bases are:

\begin{align}
\oper{U}_{A}  & = \sum\limits_{d=0}^{D-1}\left|d_{D}\right\rangle\left\langle d_{\alpha}\right|   \nonumber \\
& = \frac{1}{\sqrt{D}} \sum\limits_{d=0}^{D-1}\sum_{d^\prime=0}^{D-1} e^{{2 \pi i d^\prime (d + \alpha)}/{D}} \left|d_{D}\right\rangle\left\langle d^\prime_{D}\right|
\end{align}
and

\begin{align}
U_{B} & = \sum\limits_{d=0}^{D-1}\left|d_{D}\right\rangle\left\langle d_{\beta}\right| \nonumber \\
& = \frac{1}{\sqrt{D}} \sum\limits_{d=0}^{D-1}\sum_{d^\prime=0}^{D-1} e^{{2 \pi i d^\prime (-d + \beta)}/{D}} \left|d_{D}\right\rangle\left\langle d^\prime_{D}\right|.
\end{align}

To achieve these operators with the Talbot effect,  a combination of a phase gate \eqref{eq:ztheta} and a Talbot gate \eqref{eq:TalbotGate} can be used. Let us define the phases $\vec{\theta}=\{\theta_0,\ldots,\theta_{D-1}\}$ of the phase gate using
\begin{equation}
\theta_d(\gamma) = \frac{\pi}{4} - \frac{2 \pi \gamma d}{D} -\frac{\pi d^2}{D} ,  
\label{eq:thetad1}
\end{equation}
for even $D$ and 
\begin{equation}
\theta_d(\gamma) = \frac{\pi}{4}(D-1) - \frac{2 \pi \gamma d}{D} -\frac{\pi d^2(D+1)^2}{D} ,  
\label{eq:thetad2}
\end{equation}
for odd $D$, where $\gamma = \alpha, \beta$. 
Considering  the operation $T_{cD}Z_{\vec{\theta}}$, where again $c=1$ for odd $D$ and $c=2$ for even $D$, one can show that
\begin{equation}
\oper{T}_{cD}Z_{\vec{\theta}} = \sum_d \ket{d_D} \bra{d_\gamma} e^{-i \pi d^2/D},
\end{equation}
which is the required form for the operators $\oper{U}_A$, $\oper{U}_B$.  Thus, through application of the phase gate and the Talbot gate, it is possible to project in the bases \eqref{eq:k} and \eqref{eq:l}, allowing test of $D$-dimensional Bell inequalities using the Talbot effect.   We will illustrate our results with numerical examples in the next section.

\subsection{Numerical Example}
\label{sec:bellexample}
\begin{figure}
\includegraphics[width=8cm]{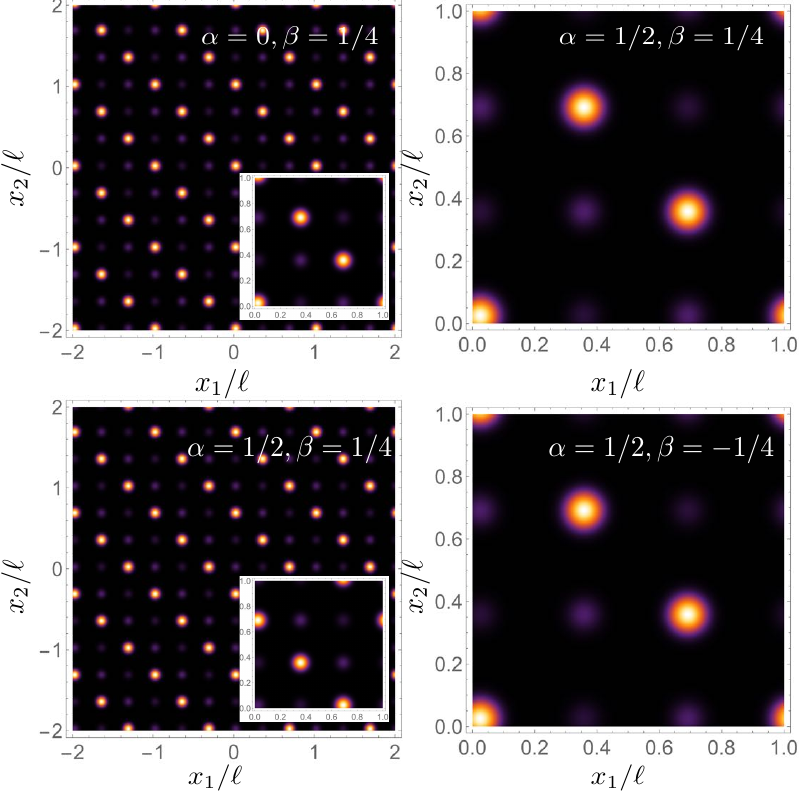}
\caption{(Color online) Plots of the joint probability distributions $P_{\alpha,\beta}(x_1,x_2)$ at Alice and Bob's detection planes for the four combination of measurements for $D=3$.  The figures on the left show several unit cells of the periodic distributions (as well as an inset of a single unit cell), while the figures on the right show a single unit cell.}
\label{fig:BellPlots3}
\end{figure}

\begin{figure}
\includegraphics[width=8cm]{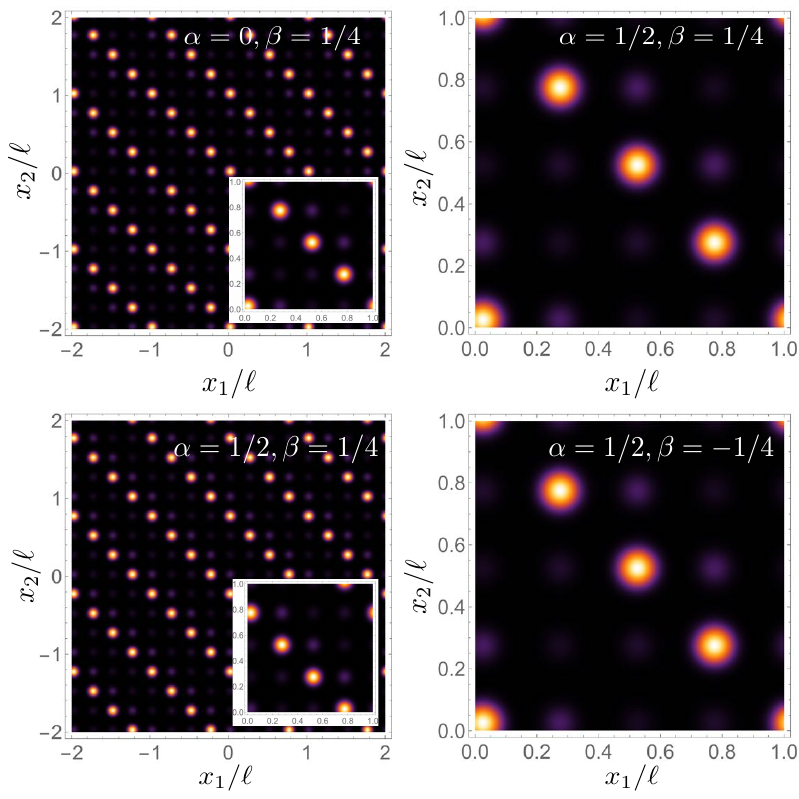}
\caption{(color online) Plots of the joint probability distributions $P_{\alpha,\beta}(x_1,x_2)$ at Alice and Bob's detection planes for the four combination of measurements for $D=4$.  The figures on the left show several unit cells of the periodic distributions (as well as an inset of a single unit cell), while the figures on the right show a single unit cell.}
\label{fig:BellPlots4}
\end{figure}

We tested the violation of the CGLMP inequalities considering that photon pairs are accurately described by the two-photon wave function \eqref{eq:PsiFinal2}, with the effective grating functions $T$ given by Gaussians.  We first applied a diagonal phase gate \eqref{eq:ztheta} with phase defined by Eq. \eqref{eq:thetad1} or \eqref{eq:thetad2}, and then propagated each grating state $T_d$ using the propagation relation \eqref{eq:propgen}.  
In Figs. \ref{fig:BellPlots3} and \ref{fig:BellPlots4} we show the joint probability distribution for the initial state as well as the four measurement combinations of the two photons for $D=3$ and $D=4$, respectively.  We use the same width and length parameters as those in Figs. \ref{fig:Carpet_Gen} b), d) and f).  The figures show the probability distribution corresponding to the entangled quantum carpets for each of the combined measurement settings.  Figures on the left show several unit cells, while figures on the right show a single unit cell of the periodic carpet distributions. 
\par
\begin{figure}
\includegraphics[width=0.48\textwidth]{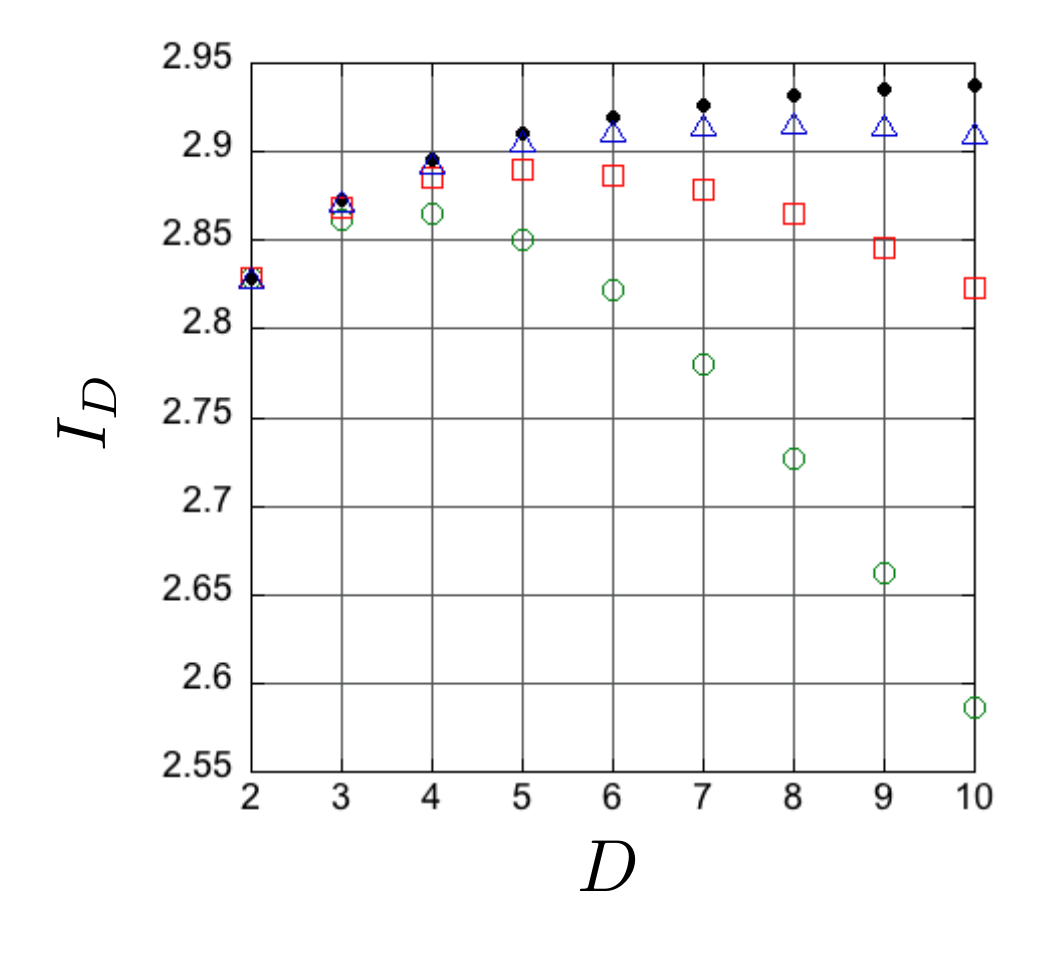}
\caption{(Color online) Numerical values of the CGLMP inequality \eqref{eq:ID} as a function of the dimension $D$.  Shown are numerical results for different amounts of initial spatial entanglement, characterized by the correlation coefficient $R$ (see text). Black solid circles correspond to maximally entangled states ($R=1$), blue triangles to $R=0.99998$, red squares to $R=0.9998$, and green circles to $R=0.998$.  }
\label{fig:ResultsSum}
\end{figure}

Figure \ref{fig:ResultsSum} shows the numerical value of the Bell parameter $I_D$ calculated for entangled quantum carpets as a function of $D$.  Shown are results for several values of initial spatial entanglement.  Black solid circles correspond to a maximally entangled state.  One can see violation grows as a function of $D$ until a maximum is reached, and that there is a maximum violation achievable for each amount of spatial entanglement.  Moreover, the maximum violation corresponds to an optimal dimension $D$ for encoding.  These results show that, for purposes of violating the CGLMP inequality, high quality spatial entanglement at the SPDC source is crucial for increasing the dimension of the entangled state.  
\section{Discussion}
All of the optical techniques used above have analogies in atom optics.  In this section, we briefly discuss the possibility of the realization of entangled quantum carpets using matter waves. 
\subsection{Possibility of a Bell test with material particles}
The observation of interference effects with material particles is one of the clearest demonstrations of the wave-like behavior of matter in quantum mechanics. In a Young's $D$-slit experiment, interference can be revealed if the width $a$ of the slits in the $D$-slit aperture and the de Broglie wavelength $\lambda_B=h/(m v)$, where $m$ is the mass and $v$ the velocity of the particles, are of the same order of magnitude. Consequently, an observation of the Talbot effect with matter waves in the near-field after a $D$-slits aperture {is not only possible, but  in the last few decades scientists have been successfully observing matter wave interference} effects with atoms and molecules with increasingly large masses \cite{hornberger12}. In this Section, we comment briefly on the possibility of performing a Bell inequality test with entangled material particles using the Talbot effect. 

Following the procedure developed in the optical context in the previous sections, the first step is to create a spatially entangled pair of material particles in an EPR state, which then can be used as starting point for implementing the scheme introduced in Sec. \ref{sec:EntCreation}. However, while the wave-like behavior of atoms and large molecules is routinely observed in state-of-the-art experiments, the creation of entanglement between such material particles is rather difficult. Here, we comment on two experimental proposals that allow for the creation of {pairs} of entangled atoms originating from Bose-Einstein condensates (BECs). The first one is based on a four-wave mixing process of two colliding BECs induced by a stimulated Raman transitions. If the collision strength is sufficiently weak, the production of entangled EPR pairs of metastable helium atoms becomes feasible \cite{Kofler}. The second proposal is based on controlled molecular Feshbach dissociation of molecules in a dilute BEC by applying weak dissociation pulses \cite{Feshbach1,Feshbach2,Feshbach3}. In both proposals the atoms can be further processed while falling in free space.

Once a spatially entangled EPR pair of two atoms has been produced, the entanglement creation scheme in Sec. \ref{sec:EntCreation} can be applied in order to transform the EPR state into an entangled Talbot state. Thus, we need to find analogs of the optical elements used to implement the $D$-slits, the gratings and the Fourier transform.  $D$-slits and gratings with an appropriate slit width and distance can be implemented using either nano-fabricated material gratings, or optical phase gratings which itself are realized by intra cavity laser fields \cite{hornberger12}. Furthermore, free propagating particles will independently undergo a Fourier transform via  propagation from the near-field to the far-field region after the respective optical elements.

At this stage we have in hand an entangled Talbot state of material particles that can in principle be used to perform a test of the Bell inequalities discussed in section \ref{sec:bell}. Thereby, the most challenging part will be to implement the phase gate (\ref{eq:ztheta}) with propagating atoms. One possibility might be to manipulate the matter wave with light pulses that themselves are controlled via a SLM in a similar manner, as discussed in \cite{Pruvost}. We leave the details of this for future investigation.
\section{Conclusion}
We have proposed a method to produce free-propagating entangled quantum carpets using spatially entangled photons produced by SPDC.  Interesting aspects of these two-photon states are: 1) the entanglement can be controlled using the initial spatial correlations of the photons, 2) $D$-dimensional entangled states can be produced, 3) quantum logic operations can be performed using free-space propagation in combination with a phase element.  We also briefly discussed the extension of these ideas to the atom optics.   We expect our results to allow for interesting investigations of quantum information tasks in $D$-dimensions. 
\begin{acknowledgements} This research was supported by the Brazilian agencies FAPERJ, CNPq, and the National Institute of Science and Technology for Quantum Information, and by the Brazil/France CAPES-COFECUB project Ph-855/15. O.J.F.  was  supported  by  the  Beatriu  de  Pin os  fellowship  (n  2014  BP-B  0219)  and  Spanish MINECO (Severo Ochoa grant SEV-2015-0522).
\end{acknowledgements}

\end{document}